\documentclass[final,5p,times,twocolumn,sort&compress]{elsarticle}

\usepackage{amssymb}
\usepackage{lipsum}
\usepackage{amsmath}
\usepackage{graphicx}
\usepackage{xcolor}
\usepackage{subcaption} 
\usepackage[ruled, lined, linesnumbered, longend]{algorithm2e}

\definecolor{webgreen}{rgb}{0,.5,0} 
\SetKwInOut{KwIn}{Input}
\SetKwInOut{KwOut}{Output}

\SetCommentSty{mycommfont}
\newcommand{\mycomment}[1]{\tcp{#1}}

\usepackage{multirow}
\usepackage{array}
\usepackage{booktabs}
\usepackage{xfrac}
\usepackage{color}

\usepackage{comment}
\usepackage{hyperref}
\hypersetup{
    colorlinks=true,
    linkcolor=blue,
    filecolor=magenta,      
    urlcolor=cyan,
    pdftitle={On-the-fly spectral unmixing based on Kalman filtering},
    }

\renewcommand{\vec}{\mathsf{vec}}
\newcommand{\unvec}{\mathsf{unvec}}
\def\tr{\mathop{\mathsf{trace}}}

\def\kron{\mathop{\mathsf{\otimes }}} 

\newcommand{\re}{\mathrm{Re}}
\newcommand{\im}{\mathrm{Im}}
\newcommand\norm[1]{\left\lVert#1\right\rVert}

\newcommand{\T}{\intercal}
\newcommand{\normal}{\mathcal{N}}

\newcommand{\bY}{\mathbf{Y}}

\newcommand{\bs}{\mathbf{s}}

\newcommand{\bsone}{\boldsymbol{1}}
\newcommand{\bszero}{\boldsymbol{0}}

\newcommand{\bsC}{\boldsymbol{C}}
\newcommand{\bsc}{\boldsymbol{c}}
\newcommand{\bsD}{\boldsymbol{D}}
\newcommand{\bsE}{\boldsymbol{E}}
\newcommand{\bsF}{\boldsymbol{F}}
\newcommand{\bse}{\boldsymbol{e}}

\newcommand{\bsG}{\boldsymbol{G}}
\newcommand{\bsy}{\boldsymbol{y}}
\newcommand{\bsY}{\boldsymbol{Y}}

\newcommand{\bsX}{\boldsymbol{X}}

\newcommand{\bsZ}{\boldsymbol{Z}}
\newcommand{\bsv}{\boldsymbol{v}}
\newcommand{\bsV}{\boldsymbol{V}}

\newcommand{\bsw}{\boldsymbol{w}}
\newcommand{\bsW}{\boldsymbol{W}}

\newcommand{\bsU}{\boldsymbol{U}}
\newcommand{\bsm}{\boldsymbol{m}}

\newcommand{\bsP}{\boldsymbol{P}}
\newcommand{\bsr}{\boldsymbol{r}}
\newcommand{\bsR}{\boldsymbol{R}}
\newcommand{\bss}{\boldsymbol{s}}
\newcommand{\bsH}{\boldsymbol{H}}
\newcommand{\bsS}{\boldsymbol{S}}

\newcommand{\bsI}{\boldsymbol{I}}
\newcommand{\bsK}{\boldsymbol{K}}
\newcommand{\bsCov}{\boldsymbol{\Sigma}}
\newcommand{\bsmu}{\boldsymbol{\mu}}
\newcommand{\bslambda}{\boldsymbol{\lambda}}

\newcommand{\htS}{\Hat{\Tilde{\bsS}}}
\newcommand{\hts}{\Hat{\Tilde{\bss}}}

\def\C{\mathbb{C}}
\def\R{\mathbb{R}}

\def\Lcal{{\mathcal{L}}}

\journal{a journal}

\begin{document}

\begin{frontmatter}

\title{On-the-fly spectral unmixing based on Kalman filtering}

\tnotetext[]{This work has been supported by the ANR IMAGIN Research Project under grant agreement ANR-21-CE29-0007.}

\author[label1]{Hugues Kouakou}
\author[label1]{José Henrique de Morais Goulart}
\author[label2]{Raffaele Vitale}
\author[label3]{Thomas Oberlin}
\author[label4]{David Rousseau}
\author[label2]{Cyril Ruckebusch}
\author[label1]{Nicolas Dobigeon}

\affiliation[label1]{organization={Université de Toulouse, IRIT/INP-ENSEEIHT},
            addressline={31071 Toulouse}, 
            country={France}}

\affiliation[label2]{organization={Université de Lille, CNRS, LASIRE},
            addressline={59000 Lille}, 
            country={France}}

\affiliation[label3]{organization={Université de Toulouse, ISAE-SUPAERO},
            addressline={31400 Toulouse}, 
            country={France}}
            
\affiliation[label4]{organization={Université d’Angers, LARIS, UMR IRHS INRAe},
            addressline={49000 Angers}, 
            country={France}}

\begin{abstract}
This work introduces an on-the-fly (i.e., online) linear unmixing method which is able to  sequentially analyze spectral data acquired  
on a spectrum-by-spectrum basis. After deriving a sequential counterpart of the conventional linear mixing model, the proposed approach recasts the linear unmixing problem into a linear state-space estimation framework. 
Under Gaussian noise and state models, the estimation of the pure spectra can be efficiently conducted by resorting to Kalman filtering. Interestingly, it is shown that this Kalman filter can operate in a lower-dimensional subspace while ensuring the nonnegativity constraint inherent to pure spectra. This dimensionality reduction allows significantly lightening the computational burden, while leveraging recent advances related to the representation of essential spectral information. The proposed method is evaluated through extensive numerical experiments conducted on synthetic and real Raman data sets. The results show that this Kalman filter-based method offers a convenient trade-off between unmixing accuracy and computational efficiency, which is crucial for operating in an on-the-fly setting. To the best of the authors' knowledge, this is the first operational method which is able to solve the spectral unmixing problem efficiently in a dynamic fashion. It also constitutes a valuable building block for benefiting from acquisition and processing frameworks recently proposed in the microscopy literature, which are motivated by practical issues such as reducing acquisition time and avoiding potential damages being inflicted to photosensitive samples. 
\end{abstract}


\end{frontmatter}

\section{Introduction}
Spectroscopic imaging, also referred to as imaging spectroscopy or hyperspectral imaging, is a widely used {measurement} technique thanks to its ability to provide rich information about sample optical properties at different wavelengths. Applications of spectroscopy arise in many domains ranging from pharmaceutics \cite{dincc2023two, wartewig2005pharmaceutical} where it can be used for checking the authenticity of drugs \cite{rebiere2018raman} to medicine \cite{ten2023analysis, bedia2021multimodal, sakudo2016near, kong2015raman}, e.g., to detect diseases \cite{kong2015raman}. 
However, a common issue inherent to hyperspectral imaging arises when the spatial resolution is above the size of pure spectral objects. This leads to mixed spectral pixels, which makes subsequent data analysis challenging. Indeed, mixed spectral pixels contain the contributions of several primary components constituting the sample under study, referred to as endmembers or pure spectra (PS) in the literature. Hence, one ubiquitous task associated to hyperspectral data consists in extracting these PS with their respective concentrations for each acquired spectral pixel. This process is called spectral unmixing (SU) \cite{Bioucas_IEEE_JSTARS_2012} or multivariate curve resolution in the field of chemometrics \cite{tauler1993multivariate}.

Since the introduction of spectral mixture analysis in the 1970s, many SU algorithms have emerged. Under a linear mixing model (LMM), estimating the PS and their respective {concentrations} in each measured spectral pixel amounts to solving a nonnegative matrix factorization (NMF) problem \cite{gillis2020nonnegative}. NMF involves decomposing a matrix into a product of two non-negative matrices. For spectral data, these two matrices contain  the PS profiles and their respective concentrations.

A well-known method for solving the SU problem is multivariate curve resolution-alternating least squares (MCR-ALS) \cite{tauler1993multivariate, tauler1995multivariate}, which formulates SU as a nonlinear least-squares problem. This algorithm is based on the simple observation that, even though SU is a (jointly) nonconvex problem, it becomes (separately) convex when either the PS  or the concentrations are kept fixed. MCR-ALS thus proceeds iteratively, by fixing one of those unknown matrices and solving the (easier) subproblem of minimizing only with respect to the other one, until some stopping criterion is met. Each subproblem is solved taking into account specific physically-driven constraints on the PS (e.g., nonnegativity, unimodality) and/or on the concentrations (e.g., positivity, sum-to-one). 

Other works have proposed to formulate SU within a Bayesian framework. Such Bayesian methods rely upon postulating some prior distributions for the PS and for the concentrations, which are carefully chosen to incorporate the problem constraints. Then, given the measured observations, the posterior distributions of the unknown quantities (PS and concentrations) are determined. In practice, these posteriors are typically intractable, hence computational methods such as Markov chain Monte Carlo sampling are used to approximate the Bayesian estimates \cite{moussaoui2006bayesian,dobigeon2009bayesian,arngren2011unmixing}.

Another class of methods, mainly motivated by geometric considerations, has also been developed. These methods can be divided into two groups, according to whether they rely or not on the assumption that at least some of the measurements are not mixed, but rather contain the sought pure spectra. Under the LMM and a sum-to-one constraint on the concentrations, the observed spectra are convex combinations of pure spectra (PS). From a geometric point of view, this means that those spectra lie within a simplex whose vertices are the PS, which are sometimes available among the collected data. A well-known algorithm which makes this assumption is  vertex component analysis (VCA) \cite{nascimento2005vertex}. VCA iteratively finds PS by projections onto random axes orthogonal to the subspace spanned by the already extracted PS, so that the next PS is set as the observation with the maximal projection coefficient in absolute value. Other algorithms that rely on the observation of PS are the successive projection algorithm (SPA) \cite{araujo2001successive}, pixel purity index (PPI) \cite{boardman1995mapping} and N-finder (N-FINDR) \cite{winter1999n}. However, this rather strong assumption is not fulfilled in many real-world scenarios. Therefore, other methods seek instead the simplex with the minimum volume enclosing the observations, without assuming the presence of observed PS. Among them, one can cite minimum volume enclosing simplex (MVES) \cite{chan2009convex}, minimum volume simplex analysis (MVSA) \cite{li2015minimum}, simplex identification via split augmented Lagrangian (SISAL) \cite{bioucas2009variable} and minimum volume constrained nonnegative matrix factorization (NMF-MVC) \cite{miao2007endmember}. 

Regardless of the used algorithm, SU performance obviously depends on the quality of the data acquisition procedure. Acquiring high-quality data comes with its own challenges: often it is quite (or even prohibitively) time-consuming, and involves long exposure to high-power excitation signals, thus potentially inflicting damages on biological samples. To address these challenges, innovative acquisition protocols have been recently developed \cite{coic2023assessment, gilet2024superpixels} to target only essential spectral pixels (ESPs), which are loosely defined as the most dissimilar spectral pixels of an image {\cite{ghaffari2019essential, ruckebusch2020perspective}}. In \cite{thurau2009convex} ESPs are found by identifying the convex hull of the collected data cloud in a subspace of reduced dimensionality which is determined via principal component analysis (PCA). Yet, PCA requires first acquiring all possible spectral pixels, which is somehow limiting. Instead, Coic \emph{et al.} have proposed to extract ES in the Fourier domain \cite{coic2023assessment}, by computing the convex hull within phasor plots \cite{scipioni2021phasor, chiang2023hyu}. This strategy has shown to be well suited to a spectrum-by-spectrum acquisition approach based on the relevance of the collected signal profiles (i.e. whether they are ESPs or not). This highlights the increasing need to be able to carry out on-the-fly (i.e., online) unmixing to update PS estimates after each new measurement. However, to the best of the authors' knowledge, none of the above SU methods allows processing spectral pixels in an on-the-fly fashion. 

To satisfy this very current need, this work presents an on-the-fly unmixing method based on the Kalman filter (KF) \cite{kalman1960new}. The estimated PS are updated through the KF each time  a new measured observation is available. To reduce the computational cost of the estimation algorithm, this KF is implemented after a Fourier-based dimensionality reduction already advocated in \cite{coic2023assessment}. Consequently, it is necessary to determine the coordinates of the estimated PS in the original data space while ensuring their nonnegativity. This task is formulated as a regression problem, which is solved by the alternating direction method of multipliers (ADMM) \cite{boyd2011distributed}. Numerical experiments conducted on synthetic and real Raman data sets demonstrates the ability of the proposed method to estimate PS on-the-fly while showing comparable results with state-of-the-art offline unmixing algorithms.

The remainder of the paper is organized as follows. Section \ref{sec:related_works} presents some related works to the general problem of source separation when the measurements may be delivered or processed sequentially. Section \ref{sec:model} recasts the conventional LMM in an on-the-fly setting, yielding the so-called on-the-fly LMM. The proposed algorithm to estimate the parameter of this on-the-fly LMM model is described in Section \ref{sec:algo}. Experimental protocols relying on synthetic and real data sets are presented in Section \ref{sec:experiments}. Results and discussions are reported in Section \ref{sec:discussion}, while Section \ref{sec:conclusions} concludes the paper.

\section{On-the-fly spectral unmixing: related work}\label{sec:related_works}
The problem of on-the-fly SU tackled in this paper can be framed as a source separation task that should be achieved along the acquisition process, i.e., for processing a data stream along its delivery by a real-time measurement system. Online\footnote{The term ``on-the-fly'' used in this paper refers to the ability of updating the PS estimate once a new measurement or a batch of measurements is available. This is rather referred to as ``online'' in the signal processing literature. Both terms are used indistinctly from now on.} source separation has in particular been applied with the objective of signal denoising, whose aim is to separate noise from the signals of interest measured in real-time, e.g., to remove artefacts originating from eye movements when measuring electroencephalogram signals \cite{shayegh2006real}. Several online algorithms based on independent component analysis (ICA) have been proposed to solve this problem, see  \cite{jafarifarmand2017real, cichocki1996robust}. ICA is based on the assumption that measurements are linear combinations of statistically independent sources. Although the linearity hypothesis is assumed in the context of this work, the assumption of statistical independence between the sources (the concentrations) is generally not fulfilled.

Source separation has also found applications in Gamma-ray spectroscopy for radionuclide detection \cite{sepulcre2013sparse,kirkpatrick2009poisson, malfrait2023online}. The authors in \cite{malfrait2023online} recently proposed a method for performing this task in real time using a series of short-term measurements on the aerosol sample being analyzed. Although their approach is built upon the LMM, it includes a prior imposing that the concentrations of two consecutively processed observations must be close, which is not indispensable with the setting considered in this work.
Indeed, in our context the spectral pixels acquired between two successive measurements can be highly dissimilar, resulting in highly variable concentrations.

Online NMF and online dictionary learning (DL) define another class of methods for online source separation, where the basis vectors (that would stand for the PS in the context of SU) are updated after each new observation is recorded. Several online NMF and DL algorithms exist in the literature, including the ones described in \cite{zhao2016online,lefevre2011online,mairal2010online,guan2012online}. These algorithms update the basis vectors from  matrices containing information on previously estimated mixing coefficients (i.e., concentrations) and past observations, which are thus sequentially updated and stored. The algorithm proposed in \cite{zhao2016online} can be interpreted as an online counterpart of the one previously introduced in \cite{Fevotte_IEEE_Trans_IP_2015} as it complements the conventional NMF objective function with an additional sparsity-inducing regularization accounting for any deviations (i.e., outlier) from the conventional bilinear model generally adopted to describe spectral mixtures. In \cite{lefevre2011online}, the authors depart from the minimization of a standard quadratic cost to derive online instances of Itakura-Saito NMF \cite{fevotte2009nonnegative} well suited to analyze audio signals. The seminal work of Mairal \emph{et al.} \cite{mairal2010online} aims at adaptively learning a basis set from a data stream to perform sparse coding. Their optimization procedure is further improved in \cite{zhao2016online} to reach a better algorithmic convergence rate based on a robust stochastic approximation of the objective function, at the price of generating random samples over the probabilistic subspace spanned by the observations. Unfortunately, in their canonical implementations, all these methods do not perform any dimensionality reduction of the measurements before processing, which poses a complexity issue when dealing with high dimensional hyperspectral data. Moreover, the robust and non-quadratic cost functions considered in \cite{lefevre2011online} and \cite{fevotte2009nonnegative} significantly deviate from the conventional bilinear model. More interestingly, the computationally lightest instance of online DL introduced in \cite{mairal2010online} shares some strong similarities with the approach described in the sequel of this paper.~We will return to this point later in Section \ref{subsec: subsec3}. However, it turns out that the online DL scheme of \cite{mairal2010online} performs poorly in the particular context of on-the-fly SU targeted in this work due to an inappropriate update rule, as we will discuss ahead. 

Finally, it is worth mentioning that online unmixing has also received attention in recent years to process multi-temporal hyperspectral images, i.e., images acquired over the same scene  at different time instants. In this context, online SU aims at  estimating the PS as well as the concentrations each time a new image is acquired in order to identify possible variations (in terms of PS and concentrations) in the scene \cite{henrot2016dynamical, Thouvenin_IEEE_Trans_CI_2018, Thouvenin_IEEE_Trans_GRS_2019, nus2020admm}. These methods are clearly not adapted to the context of SU of data delivered according to a spectrum-by-spectrum acquisition protocol, as they require the registration of the entire hyperspectral image  at each time step.

\section{On-the-fly formulation of LMM}\label{sec:model}

\subsection{Linear mixing model}
This work addresses the problem of on-the-fly SU under the LMM, which postulates that each measured spectrum, that is to say each observation, is a linear combination of some unknown pure spectra (PS). The PS and their concentrations in each  measured spectrum are supposed nonnegative. Moreover, this work considers also a sum-to-one (closure) constraint imposed on the concentrations, as it is commonly done in SU to ease their interpretation as mixing proportions. It is worth noting that this constraint has the great advantage of removing the scale ambiguity inherent to linear unmixing and can always be satisfied by the data after an appropriate normalization \cite{omidikia2018closure}. Assuming that the image to be acquired is composed of $N$ measured spectral pixels, after a standard unfolding procedure, the LMM writes
\begin{equation} \label{eq1}
    \bsY = \bsC\bsS^\T + \bsE,
\end{equation}
where \(\bsY \in \R_{+}^{N \times L}\) contains the full set of $N$ $L$-dimensional spectra, \(\bsC \in \R_{+}^{N \times K}\)  is the concentration matrix where the $n$th row contains the $K$ concentrations associated to the $n$th pixel, \(\bsS \in \R_{+}^{L\times K}\) contains the $K$ unknown pure spectra  and \(\bsE \in \R^{N \times L}\) is an additive white Gaussian noise matrix which models measurement noise and/or mismatches with respect to the LMM. The number of wavelengths or wavenumbers is denoted $L$, and the number $K$ of PS is assumed to be known a priori. Moreover, the matrices \(\bsC \) and \(\bsS\) are nonnegative, i.e., they are composed of nonnegative elements, and the sum-to-one constraint is imposed on the concentrations as \(\bsC\bsone_{K}= \bsone_{N} \),  where $\bsone_{M}$ denotes the $M$-dimensional vector of ones.

\subsection{On-the-fly linear mixing model}

In an on-the-fly setting, the spectral pixels composing the full scene are acquired sequentially and should be processed on a spectrum-by-spectrum basis. Without loss of generality but to ease the presentation, one can assume that the spectral pixels are arranged in the matrix $\bsY=[\bsy_1, \ldots, \bsy_{N} ]^\T$ according to their order of acquisition, i.e., the vector \( \bsy_t\in \R_{+}^L \) defining  the $t$th row of $\bsY$ corresponds to the $t$th acquired spectrum ($t=1,\ldots,N$). Given a current estimate \( \Hat{\bsS}_{t-1}\) of the PS recovered from the set of $(t-1)$ acquired spectra \( \bsY_{\leq t-1} = [\bsy_1, \ldots, \bsy_{t-1} ]^\T\), one seeks to update it as soon as a new spectral pixel \( \bsy_t \) is acquired such that \( \Hat{\bsS}_t\) tends to the true (but unknown) PS matrix \( \bsS \). A possible approach is to use any offline SU method each time a new spectral pixel \( \bsy_t \) is available.  Doing so will give an estimate \( \Hat{\bsS}_t\) of the PS based on the spectral pixels $\bsY_{\leq t}$ acquired so far, such that 
\( \bsY_{\leq t} \approx \bsC_{\leq t} \Hat{\bsS}^\T_t\) with \( \bsC_{\leq t} = [\bsc_1, \ldots, \bsc_{t} ]^\T\) and $\bsc_t \in \R_{+}^K$. Yet, this scheme uses all available spectra and recomputes an estimate at each time instant ``from scratch,'' which is not suitable for an on-the-fly approach because of the computational burden. This naive strategy will be considered in Section \ref{sec:experiments} when comparing the proposed method to conventional unmixing algorithms. A question that arises is thus how to efficiently compute the PS estimate at instant $t$ based on the  information already available, i.e., exploiting the current estimate at instant $t-1$. One natural way of tackling this challenge is to define the PS matrix \( \bsS_t \) associated with the set of $t$ measurements $\bsY_{\leq t} $ as a random (matrix-valued) variable evolving according to \( \bsS_t = f_t(\bsS_{t-1}) + \bsV_t \), for some suitable function \(f_t(\cdot)\) and an appropriately defined random variable \(\bsV_t \). According to this formulation, an estimate of \( \bsS_{t-1}\) can be leveraged as a prior information for estimating \(\bsS_t \), together with assumptions on \( f_t(\cdot)\) and on \( \bsV_t \).

In this paper, we model the evolution of \( \bsS_t\) as a random walk without drift, i.e., $f_t(\cdot)$ is assumed to be the identity function and \( \bsV_t \) is a random matrix with zero-mean entries. This implies that \( \bsS_t\) is distributed around \( \bsS_{t-1}\). More specifically, one adopts the following on-the-fly formulation of LMM
\begin{eqnarray}
    \bsS_t &= \bsS_{t-1} + \bsV_t, \label{eq2}\\
    \bsy_t &= \bsS_t\bsc_t + \bse_t, \label{eq3}
\end{eqnarray}
where $ (\bsV_{t})_{ij} \overset{\text{i.i.d.}}{\sim} \normal(0, \sigma_v^2)$, $\bse_t \sim \normal(\bszero, \sigma_e^2 \, \bsI_L)$ and \(\bsI_L\) is the identity matrix of size $L$. The variance $\sigma_v^2$ can be interpreted as the  uncertainty level associated to the PS estimate \(\bsS_t \) with respect to the previous estimate \( \bsS_{t-1} \) as a new spectrum \(\bsy_t \) arrives. The on-the-fly LMM defined by Eq.~\eqref{eq2}--\eqref{eq3} states that the spectrum \(\bsy_{t}\) depends on the PS matrix \( \bsS_t \) according to the conventional (static) LMM, and this PS matrix evolves according to a recursion. On-the-fly unmixing can then be formulated as the updating of a previous PS estimate \( \hat{\bsS}_{t-1} \) given a new acquired spectral pixel$\bsy_t$. This can be achieved within a Bayesian framework by deriving  the (conditional) probability density function (pdf) of \( \bsS_t\) given \( \bsy_1, \ldots, \bsy_t\). This approach is detailed in the next section.

\section{The proposed on-the-fly unmixing algorithm}\label{sec:algo}
\subsection{Sequential estimation of pure spectra}
\label{subsec: subsec1}
For the sake of simplicity, this subsection assumes  that the concentrations \( \bsc_t \in \R_{+}^K\) are known; their estimation will be addressed later in Section \ref{subsec:concentrations}. Moreover, the proposed on-the-fly LMM is vectorized by stacking the columns of \(\bsS_t\) and \(\bsV_t \) after applying the $\vec(\cdot)$ operator. By using the identity  $\vec{(\bsS_t\bsc_t)}= (\bsc_t^\T \kron \bsI_L) \vec{(\bsS_t)} = \bsH_t\bss_t$ where $\kron$ denotes the Kronecker product,  the on-the-fly LMM defined as in \eqref{eq2} and \eqref{eq3} becomes
\begin{eqnarray}
    \bss_t &= \bss_{t-1} + \bsv_t, \label{eq5}\\
    \bsy_t &= \bsH_t\bss_t + \bse_t, \label{eq6}
\end{eqnarray}
where $\bss_t \in \R^{LK}$ (resp., $\bsv_t$) is the vectorized counterpart of   $\bsS_t$ (resp., $\bsV_t$) and $\bsH_t = \bsc_t^\T \kron \bsI_L \in \R_{+}^{L\times LK}$. The posterior distribution of the unknown PS $\bss_t$ writes
\[
 p(\bss_t \rvert \bsY_{\leq t} ) 
            \propto p(\bsy_t \rvert \bss_t) \, p(\bss_t \rvert \bsY_{\leq t-1}).
\]
Given the Gaussian nature of the noise term $\bsv_t$ and the error term $\bse_t$, this posterior is also a Gaussian distribution \(\normal(\bsmu_t, \bsCov_t)\), thus fully defined by its mean $\bsmu_t$ and covariance matrix $\bsCov_t$. After each new measurement $\bsy_t$ is recorded, these parameters $(\bsmu_{t},\bsCov_{t})$ can be recursively and explicitly computed from the previous parameters $(\bsmu_{t-1},\bsCov_{t-1})$ according to updating rules specified by the Kalman filter (KF), as detailed in Algo.~\ref{alg:KF-Algorithm}.  More precisely, the KF updates the conditional mean $\bsmu_t$ and covariance matrix $\bsCov_t$ at each time instant $t$ following two steps, namely prediction and update. The prediction step calculates the predicted covariance matrix \( \bsCov_{t-\sfrac{1}{2}}\) of the distribution \(p(\bss_t \rvert \bsY_{\leq t-1} ) = \normal(\bss_t; \bsmu_{t-1}, \bsCov_{t-\sfrac{1}{2}})\). Once a new spectrum \( \bsy_t\) is available, the mean and covariance matrix are updated. This KF scheme is known to be optimal in the mean-square-error sense \cite{kalman1960new}. Besides, at each time instant $t$ of the KF (i.e., after each new measurement is collected), it allows a Bayesian estimator to be computed. For example, the minimum mean-square-error (MMSE) estimate boils down to the posterior mean $\bsmu_{t}$.

\begin{algorithm}
\caption{$\mathsf{KF{-}Update}$}\label{alg:KF-Algorithm}
\SetKwInOut{KwIn}{Input}
\SetKwInOut{KwOut}{Output}
        \KwIn {Newly acquired spectrum $\bsy_t$, posterior mean $\bsmu_{t-1}$ and covariance matrix $\bsCov_{t-1}$ at time instant $(t-1)$, observation matrix $\bsH_t$, uncertainty level $\sigma_v^2$, model noise variance $\sigma_e^2$} 
       \mycomment{  Predict the prior covariance matrix }
        \(  \bsCov_{t-\sfrac{1}{2}} \gets \bsCov_{t-1} + \sigma_v^2\bsI_{KL}   \)\\
        \mycomment{Compute the innovation}
         \(  \bsr_t \gets  \bsy_t - \bsH_t\bsmu_{t-1} \)\\
         \mycomment{Compute the innovation covariance}
        \( \bsZ_t \gets \bsH_t\bsCov_{t-\sfrac{1}{2}}\bsH_t^\T + \sigma_e^2\bsI_L  \)\\
        \mycomment{Compute the optimal gain}
        \( \bsK_t \gets \bsCov_{t-\sfrac{1}{2}}\bsH_t^\T\bsZ_t^{-1}   \) \label{line:Kalman_gain}\\
        \mycomment{Update the posterior mean}
        \( \bsmu_t  \gets  \bsmu_{t-1}  + \bsK_t\bsr_t  \)\\ 
        \mycomment{Update the posterior covariance matrix}
        \( \bsCov_t \gets (\bsI_L - \bsK_t\bsH_t)\bsCov_{t-\sfrac{1}{2}} \)\\
    \KwOut{Posterior mean  \( \bsmu_t \) and covariance matrix  \(  \bsCov_t\) at time instant $t$.}
    \end{algorithm}

At this stage of the estimation process, two important shortcomings should be discussed. First, in its canonical implementation, this KF-based unmixing procedure does not ensure the nonnegativity of the PS. Second, in practice, the steps of this algorithm can be quite costly, as the spectra acquired by hyperspectral cameras are high-dimensional, typically featuring hundreds or thousands of entries. Thus, it is imperative to perform dimensionality reduction before executing the KF to make it feasible for real-time SU. The next section shows that this estimation procedure can be carefully adapted to tackle these two challenges, i.e., reducing the computational cost via a dimensionality reduction while ensuring the nonnegativity constraint inherent to the PS.

\subsection{Constrained estimation in a low-dimensional subspace}
\label{subsec: subsec2}
As stated in the introduction, the most popular approach to perform dimensionality reduction consists in conducting PCA. However, this strategy is not suitable for the particular setting considered in this work, i.e,  a sequential (spectrum-by-spectrum) acquisition process combined with on-the-fly data processing. Indeed, identifying the principal components would require to  acquire  first the whole set of spectra $\bY$. Instead, this work proposes to capitalize on the study conducted in \cite{coic2023assessment}, which effectively demonstrated the suitability of the Fourier domain for ESPs identification. The central idea of this approach lies on the fact that the discrete Fourier transform (DFT) can be computed on a spectrum-by-spectrum basis, unlike PCA, which needs all the spectra at once. Moreover, it is worth noting two  interesting features of the DFT in the context of on-the-fly SU. Firstly, it is a linear transformation; as a consequence, the linearity of the LMM is preserved. Secondly, the use of DFT is compatible with real-time data processing thanks to the computational efficiency of the fast Fourier transform (FFT) algorithm.

As in \cite{coic2023assessment}, the lower-dimensional representation $\Tilde{\bsy}_t \in  \R^{2M}$ (with $M \ll L$) associated with the spectrum \(\bsy_t \in  \R_+^L \) is defined through the linear dimensionality reduction operator $\mathsf{DR}: \mathbb{R}_+^L \rightarrow \mathbb{R}^{2M}$ as
\begin{equation}
    \Tilde{\bsy}_t = \mathsf{DR}\left({\bsy}_t\right) = \begin{bmatrix}
\re{(\bsF^\T\bsy_t)}\\
\im{(\bsF^\T\bsy_t)}
\end{bmatrix}.\label{eq8}
\end{equation}
The lower dimensional representation $\Tilde{\bsy}_t$ contains the concatenation of the real and imaginary parts of the DFT coefficients for a set of $M$ frequencies computed from the truncated DFT matrix \( \bsF \in \C^{L\times M}\). The $M$ columns of \( \bsF \) are chosen such that $\Tilde{\bsy}_t$ contains sufficient energy with respect to the total mean energy of $\bsy_t$, i.e., \( \mathrm{E}\left[\norm{\Tilde{\bsy}_t}^2_2\right] \geq \tfrac{\eta}{100} \mathrm{E}\left[\norm{\bsy_t}^2_2 \right] \) for some chosen \( \eta \in (0,100]\).

Thanks to the linearity of the DFT, the on-the-fly LMM \eqref{eq5}-\eqref{eq6} can be rewritten in the lower-dimensional Fourier subspace as
\begin{eqnarray}
    \Tilde{\bss}_t &= \Tilde{\bss}_{t-1} + \Tilde{\bsv}_t, \label{eq9}\\
    \Tilde{\bsy}_t &= \bsH_t\Tilde{\bss}_t + \Tilde{\bse}_t, \label{eq10}
\end{eqnarray}
where $\Tilde{\cdot}$ refers to the corresponding quantity defined in the subspace and $\bsH_t = \bsc_t^\T \kron \bsI_{2M}$. In Algo.~\ref{alg:KF-Algorithm}, which remains unchanged, the KF can then be directly applied to the lower-dimensional representation \( \Tilde{\bsy}_t \) instead of \( \bsy_t \).

We turn now to the second challenge to tackle, i.e., ensuring the nonnegativity constraint on the PS estimate. Enforcing this constraint to the PS \( {\bss}_t \) expressed in the original data space would be rather straightforward, since it can be achieved by a simple hard-thresholding of the KF output. Unfortunately, conducting the sequential estimation in a lower-dimensional subspace makes this constraint more difficult to handle. Indeed, the KF operates in the low-dimensional subspace to track the lower-dimensional representation $\Tilde{\bss}_t $ of the PS.  Conversely, the constraint applies to the PS ${\bss}_t $ defined in the original data space. To overcome this issue, the proposed strategy  reformulates the constrained estimation problem as a constrained regression, as detailed in what follows. To ease the presentation, we go back to the non-vectorized notation of the PS estimate at time instant $t$, i.e., we denote the lower-dimensional estimate recovered by the KF at time instant $t$ as \(\htS_t = \unvec(\hts_t)\). This current estimate is supposed to be a rather good approximation of the true PS expressed in the lower-dimensional subspace, i.e., for $j\leq t$ it holds
\begin{equation}
    \Tilde{\bsy}_j \approx \hat{\Tilde{\bsS}}_t \bsc_j.
\end{equation}
Reciprocally, it is also legitimate to state that the lower-dimensional representations of the PS can be expressed as linear combinations of some lower-dimensional representations of the measured spectral pixels, i.e., there exists a regression matrix $\bsR \in \mathbb{R}^{P \times K}$ such that
\begin{equation}
    \htS_t \approx \Tilde{\bsY}_{1:P}\bsR_t
\end{equation}
where the $P$ columns of $\Tilde{\bsY}_{1:P} \in \mathbb{R}^{2M \times P}$ contains the $P$ lower-dimensional representations $\Tilde{\bsy}_1,\ldots,\Tilde{\bsy}_P$ of the measured spectra ${\bsy}_1,\ldots,{\bsy}_P$ (with $K \leq P\leq t)$. The core idea consists in translating this regression into the original data space while ensuring the nonnegativity of the recovered PS. More precisely, the nonnegative constrained PS estimate in the data space denoted $\hat{\bsS}_t^{({+})}$ is assumed to be expressed as
\begin{equation} \label{eq:PS_nonnegative}
    \hat{\bsS}_t^{({+})} = {\bsY}_{1:P}\hat{\bsR}_t
\end{equation}
where the regression matrix $\hat{\bsR}_t$ is defined as the solution of the constrained regression problem
\begin{equation}
    \hat{\bsR}_t = \operatornamewithlimits{argmin}_{\bsR_t} \norm {\Tilde{\bsY}_{1:P}\bsR_t - \htS_t}^2_\text{F} \quad    \textrm{s.~t.} \quad  \bsY_{1:P}\bsR_t \geq \bszero.
    \label{eq11}
\end{equation}
The nonnegativity of the PS estimate defined in \eqref{eq:PS_nonnegative} is directly ensured by the nonnegativity constraint included in the regression problem \eqref{eq11}. Note that incorporating this constraint into the original measurement space overcomes the non-invertibility of the dimensionality reduction induced by \eqref{eq8}. Indeed, it only relies on the straightforward one-to-one mapping between the set of $P$ spectra expressed in the original space and their representations  with respect to the chosen lower-dimensional subspace (DFT) basis, namely, ${\bsY}_{1:P}$ and ${\Tilde{\bsY}}_{1:P}$. We propose to solve the minimization problem in (\ref{eq11}) by means of the alternating direction method of multipliers (ADMM)  \cite{boyd2011distributed}. Interested readers are invited to refer to \ref{sec: appendix1} for further details, and in particular for an algorithmic sketch of the regression procedure. 

Throughout Sections \ref{subsec: subsec1} and \ref{subsec: subsec2}, the concentrations \(\bsc_t\) ($t\geq 1$), or equivalently the observation model matrices $\bsH_t$, have been assumed known, which is obviously not the case in practice. The next subsection discusses their estimation along the KF iterations (or time instants).

\subsection{Estimation of concentrations}\label{subsec:concentrations}
Once a spectral pixel \(\bsy_t \) is observed at a given instant \( t\), the associated concentration vector \(\bsc_t \) defining the observation matrix $\bsH_t$ is required to update the PS estimate. Several strategies can be envisaged to infer this quantity. Driven by the constraint of computational efficiency compatible with on-the-fly processing, this work proposes to follow the same strategy advocated in recent works dedicated to online NMF \cite{zhao2016online, guan2012online} and online DL \cite{mairal2010online}. More precisely, an estimate of \( \bsc_t\) can be computed from the current constrained estimate  \( \hat{\bsS}^{(+)}_{t-1} \) of the PS defined in \eqref{eq:PS_nonnegative} by solving the constrained regression problem
\begin{equation}
    \min_{\bsc} \norm { {\bsy}_t - \hat{\bsS}^{(+)}_{t-1} \bsc}^2_2 \quad  \textrm{s.~t.} \quad   \bsc \geq \bszero \quad  \textrm{and} \quad \bsc^\T\bsone_{K}=1.
    \label{eq13}
\end{equation}
Various efficient off-the-shelf algorithms can be considered to solve this constrained minimization problem, such as the sparse unmixing by variable splitting and augmented Lagrangian (SUnSAL) algorithm \cite{bioucas2010alternating}.

\subsection{Overview of the proposed algorithm}

\begin{algorithm}[t]
\caption{Proposed {KF-OSU} algorithm for on-the-fly spectral unmixing}
\label{alg:Method}
\SetKwInOut{KwIn}{Input}
\SetKwInOut{KwOut}{Output}
    \KwIn{Number $K$ of PS, number $P$ of regressors, $P$ measured spectra $\bsY_{1:P}$, model noise variance $\sigma^2_{e}$, uncertainty level $\sigma^2_v$}
    \SetKwInOut{KwIn}{Initialization}
    \KwIn {Constrained PS estimate $\bsS^{(+)}_p$}
    \( t\gets P\) \\
    \mycomment{Perform dimensionality reduction (see \eqref{eq8})}
    \(  \Tilde{\bsS}^{(+)}_t \gets \mathsf{DR}\left(\bsS^{(+)}_t\right) \)\\ 
     \mycomment {{Initialize the posterior distribution parameters}}
    \(  \Tilde{\bsmu}_t \gets \vec\left(\tilde{\bsS}^{(+)}_t\right) \)\\
    \( \bsCov_t \gets \sigma^2_v\bsI_{2KM}\)\\
    \mycomment {{ Update the PS estimate after each acquisition}}
    \Repeat{\textnormal{end of acquisition}}{
        \(t \gets t+1 \)\\ 
        \mycomment{{Estimate mixing concentrations $\bsc_t$} by solving (\ref{eq13})}
        \(\bsc_t \gets \mathsf{SUnSAL}\left({\bsy}_t,\bsS^{(+)}_{t-1}\right) \)\\
        $\bsH_t \gets \bsc_t^\T \kron \bsI_{2M}$ \\ 
        \mycomment{Perform dimensionality reduction (see \eqref{eq8})}
        \(  \Tilde{\bsy}_t \gets \mathsf{DR}(\bsy_t) \)\\
        \mycomment{{Update the posterior distribution parameters (Algo.~\ref{alg:KF-Algorithm})}}
        \(  \left(\Tilde{\bsmu}_t, \bsCov_t\right) \gets \mathsf{KF{-}Update}\left(\Tilde{\bsy}_t, \Tilde{\bsmu}_{t-1}, \bsCov_{t-1}, \bsH_t, \sigma_v^2, \sigma_e^2\right)  \) \label{line:Posterior_Update}\\
        \mycomment{{Update the unconstrained PS estimate in the subspace}}        
        \( \tilde{\bsS}_t  \gets \unvec(\Tilde{\bsmu}_t)\)\\
         \mycomment{{Compute the regression matrix by solving \eqref{eq11}} (Algo.~\ref{alg:ADMM})}
        \(\bsR_t  \gets \mathsf{Regression}\left(\bsY_{1:P}, \Tilde{\bsY}_{1:P}, \tilde{\bsS}_t \right) \) \label{algostep:regression}\\
        \mycomment{{Compute the constrained PS estimate in the full space}}
        \( {\bsS}^{(+)}_t \gets \bsY_{1:P}\bsR_t   \)\\                
        \mycomment{Perform dimensionality reduction (see \eqref{eq8})}
        \(  \Tilde{\bsS}^{(+)}_t \gets \mathsf{DR}\left(\bsS^{(+)}_t\right) \)\\ 
        \mycomment{{Update the posterior mean}}
        \(\Tilde{\bsmu}_t  \gets  \vec{\left(\tilde{\bsS}^{(+)}_t\right)} \)\\
    }
     \KwOut{The estimated PS matrix \({\bsS}^{(+)}_t\).}
    \end{algorithm}

A step-by-step description of the proposed KF-based on-the-fly SU algorithm, termed {KF-OSU}, is sketched in Algo.~\ref{alg:Method}. To lighten the notations, the estimate symbols $\hat{\cdot}$ over the various quantities of interest have been omitted without any ambiguity. The constrained PS estimate \(\bsS^{(+)}_p\) can be initialized as the output of  any offline unmixing algorithm applied to the first $P$ measured spectra. 

\subsection{Connection to related methods} 
\label{subsec: subsec3}
The core step of the proposed KF-OSU algorithm lies in the update of the PS estimate after each newly acquired measurement $\bsy_t$. In this work, this update is performed within a fully Bayesian framework by Kalman filtering (line \ref{line:Posterior_Update} in Algo.~\ref{alg:Method}). Conceptually, one may think of other updating strategies. This paragraph discusses these alternatives and relates them to the currently implemented KF-based updating rule.

As a preliminary, it should be noticed that the KF-based updating rule fully detailed in Algo. \ref{alg:KF-Algorithm} can be compactly rewritten as
\begin{equation}\label{eq:KF_update}
    \bs_t^{\textrm{KF}} = \bs_{t-1}^{\textrm{KF}} + \bsK_t\left(\bsy_t-\bsH_t \bs_{t-1}^{\textrm{KF}}\right)
\end{equation}
where $\bsK_t$ is the so-called optimal Kalman gain (see line \ref{line:Kalman_gain} of Algo.~\ref{alg:KF-Algorithm}) resulting from the assumptions of linear-Gaussian models for the state \eqref{eq5} and the observation \eqref{eq6}. It mainly consists in first computing a residual $r_t^{\textrm{KF}} = \bsy_t-\bsH_t \bs_{t-1}^{\textrm{KF}}$, also referred to as innovation within the KF framework. This residual $r_t^{\textrm{KF}}$ acts as a drift, weighted by the gain matrix $\bsK_t$, to  correct the previous estimate $\bs_{t-1}^{\textrm{KF}}$.

Then, because of the simple state evolution model \eqref{eq5}, it can be shown that there is a clear connection with the updating rule that would result from a recursive least square (RLS) filtering \cite{haykin2002adaptive}. RLS aims at solving at each time instant $t$ the weighted least-squares minimization problem
\begin{equation}\label{eq:RLS_problem}
\operatornamewithlimits{min}_{\bss} \sum_{i=1}^{t} \lambda^{t-i}\norm{ {\bsy}_i - \bsH_i\bss}^2_2,
\end{equation}
where $\lambda \in (0,1)$ is a forgetting factor. This leads to an RLS update rule of the form
\begin{equation}\label{eq:RLS_update}
{\bss}^{\textrm{RLS}}_t = {\bss}^{\textrm{RLS}}_{t-1} +  \bsG_{t}(\lambda) \left( {\bsy}_t - \bsH_t{{\bss}}^{\textrm{RLS}}_{t-1} \right),
\end{equation}
where $\bsG_{t}(\lambda)$ can be interpreted as a RLS gain. Comparing \eqref{eq:KF_update} and \eqref{eq:RLS_update}, it clearly appears that KF and RLS differs by the definition of the gain applied to the correction drift.

Besides, as already discussed in Section \ref{sec:related_works}, on-the-fly SU can also be  cast into online NMF or online DL frameworks. In particular, the most popular strategy to perform online DL is certainly the stochastic optimization algorithm introduced by Mairal \emph{et al.} in \cite{mairal2010online}. In its canonical implementation, updating the dictionary atoms (associated with the PS in the context of SU) is achieved by solving the optimization problem
\begin{equation}\label{eq:DL_problem}
\operatornamewithlimits{min}_{\bsS} \sum_{i=1}^{t} \norm{ {\bsy}_i - \bsS \bsc_i}^2_2. 
\end{equation}
Following the vectorization procedure described in Section \ref{subsec: subsec1}, it can be shown that the original DL updating rule to solve \eqref{eq:DL_problem} and given in \cite{mairal2010online} can be rewritten as
\begin{equation}\label{eq:DL_update}
{\bss}^{\textrm{DL}}_t = {\bss}^{\textrm{DL}}_{t-1} +  \bsD_{t} \left( {\bsy}_t - \bsH_t{{\bss}}^{\textrm{DL}}_{t-1} \right)
\end{equation}
with
\begin{equation}
 \bsD_t = \left( \left(\sum_{i=1}^{t} \bsc_i\bsc_i^\T \right)^{-1}\bsc_t \right) \kron \bsI_{L}. 
\end{equation}
Again, by interpreting \(\bsD_t\)  as a gain, the link between this DL update and the KF update \eqref{eq:KF_update} is clear.

Numerical experiments conducted on synthetic data sets empirically showed that replacing the KF update (line \ref{line:Posterior_Update} in Algo.~\ref{alg:Method}) with one of these two alternative updating rules \eqref{eq:RLS_update} or \eqref{eq:DL_update} did not improve or even severely degraded the PS estimation. This may due to an unsuitable definition of the gain matrices related to these alternative methods. As a consequence, they will not be considered further in the sequel of this paper.

\section{Numerical experiments}\label{sec:experiments}

This section describes the numerical experiments conducted to assess the performance of the proposed method on synthetic and real Raman data sets. It describes two synthetic data sets and a real data set used in the experiments, introduces two simulation protocols and describes the algorithms against which KF-OSU is compared. In addition to their ability to properly recover the sought PS, these methods will be compared with respect to their computational burden (i.e, algorithmic runtimes).

\subsection{Description of the data sets}

\subsubsection{Synthetic data sets}

\paragraph{Data sets \texttt{SD1}} Two hundred data sets, each composed of 4000 spectra, have been independently generated from a mixture of $K=5$ spectra according to the linear mixing model in (\ref{eq1}). Mixture concentrations have been first randomly drawn from a Dirichlet distribution $\mathcal{D}(\boldsymbol{\alpha})$ guaranteeing their nonnegativity and their sum-to-one constraints. The Dirichlet parameter $\boldsymbol{\alpha}=\left[\alpha_1,\ldots,\alpha_K\right]$ with $\alpha_k=1$ (for all $k$), i.e., the mixing coefficients are uniformly distributed over the simplex defined by the constraints. Each generated data set contained pure spectra. Finally, the noise term $\bsE$ is drawn from a white Gaussian distribution where the noise variance $\sigma^2_{e}$ has been adjusted to achieve a signal-to-noise ratio (SNR) of $20$dB. 

\paragraph{Data sets \texttt{SD2}} The  generation procedure for this second set of data is the same as for \texttt{SD1}. However, all spectra containing more than 75\% of one of the primary spectra have been discarded, to eliminate almost pure spectra and to make the unmixing problem more challenging. 

\subsubsection{Real data set \texttt{RD}}
\begin{figure}[t]
  \centering
\includegraphics[width=0.8\linewidth]{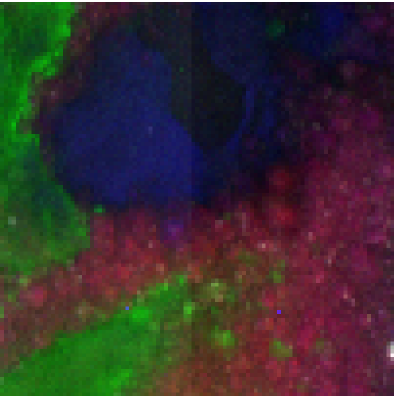}  \caption{Data set \texttt{RD}: RGB composition of the real Raman hyperspectral image used in the experiments. The red (R), green (G) and blue (B) channels correspond to Raman shifts $711.8$, $622.9$ and $723.6$ cm$^{-1}$, respectively.}
  \label{fig:fig1}
\end{figure}

A $101\times101\times343$ Raman image, depicted in Fig.~\ref{fig:fig1} has been obtained from a mixture of sodium nitrate (NaNO$_3$), calcium carbonate (CaCO$_3$) and sodium sulphate (Na$_2$SO$_4$) powders. All spectral pixels have been acquired with a LabRAMHR microspectrometer (Horiba France SAS, Palaiseau, France) using a 50{\normalsize$\times$} Olympus objective (0.75 NA). Data acquisition has been performed in the spectral range $497.7 - 901.2$ cm$^{-1}$ with an accumulation time of $5$ s per pixel (see \cite{coic2023assessment} for more details on the acquisition conditions). This real data set denoted \texttt{RD} is interesting for assessing the performance of the compared  methods as reference spectra, i.e.~the spectral signatures of the three components NaNO$_3$, CaCO$_3$ and Na$_2$SO$_4$, are available. After unfolding the acquired hyperspectral image, a $N\times L$ matrix $\bsY$ of spectra is obtained with $L=343$ and $N=10201$. A batch of 199 copies of this matrix has been generated by randomly permuting the rows. The goal is to evaluate the performance of the proposed {KF-OSU} method for different initializations and sequences of observed spectra.

\subsection{Acquisition protocols}
Two distinct acquisition protocols have been simulated. For the first protocol denoted \texttt{P1}, the available spectral pixels are all kept and no procedure for selecting essential spectral pixels and no preferred order of measurement (observation) has been implemented. The second protocol, denoted \texttt{P2}, is inspired by the principle of essential spectral pixels \cite{coic2023assessment, gilet2024superpixels}.
  More precisely, the data sets generated under protocol \texttt{P2} consist of only essential spectral pixels extracted from the data resulting from protocol \texttt{P1}. These selected pixels are subsequently arranged in a specific order to maximize the spectral diversity (i.e., dissimilarity) between consecutive observed spectra. This order relies on an iterative \emph{peeling} procedure of the data set conducted by identifying successive convex hulls, in a fashion similar to the strategy advocated in \cite{ahmad2023weighted}. The procedure is fully described in \ref{sec: appendix2}. The main motivation behind protocol \texttt{P2} is to illustrate the impact of the order in which measurements are, carried out and then made available for analysis by the on-the-fly unmixing algorithm. In particular, it will help to show  that the convergence  of KF-OSU can be significantly accelerated by providing the most informative measured spectra at an early stage of the estimation procedure.

\begin{figure*}[t]
  \centering
\includegraphics[width=1\linewidth]{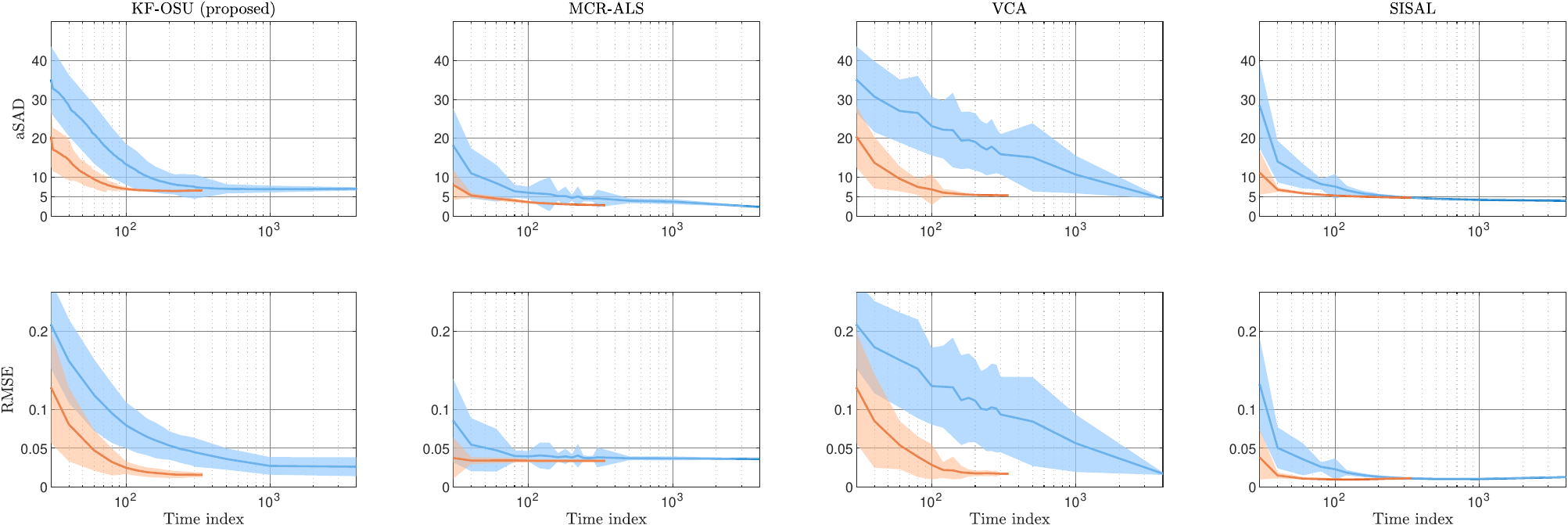}  \caption{Synthetic data set \texttt{SD1} -- Performance of the compared algorithms in terms of aSAD ($1$st row) and RMSE ($2$nd row) as functions of the time index. Blue and orange lines refer to the acquisition protocols \texttt{P1} and \texttt{P2}, respectively. The results have been averaged over $200$ data sets and the shaded areas correspond to one standard deviation.}
  \label{fig:fig3}
\end{figure*}

\begin{figure*}[t]
  \centering
\includegraphics[width=1\linewidth]{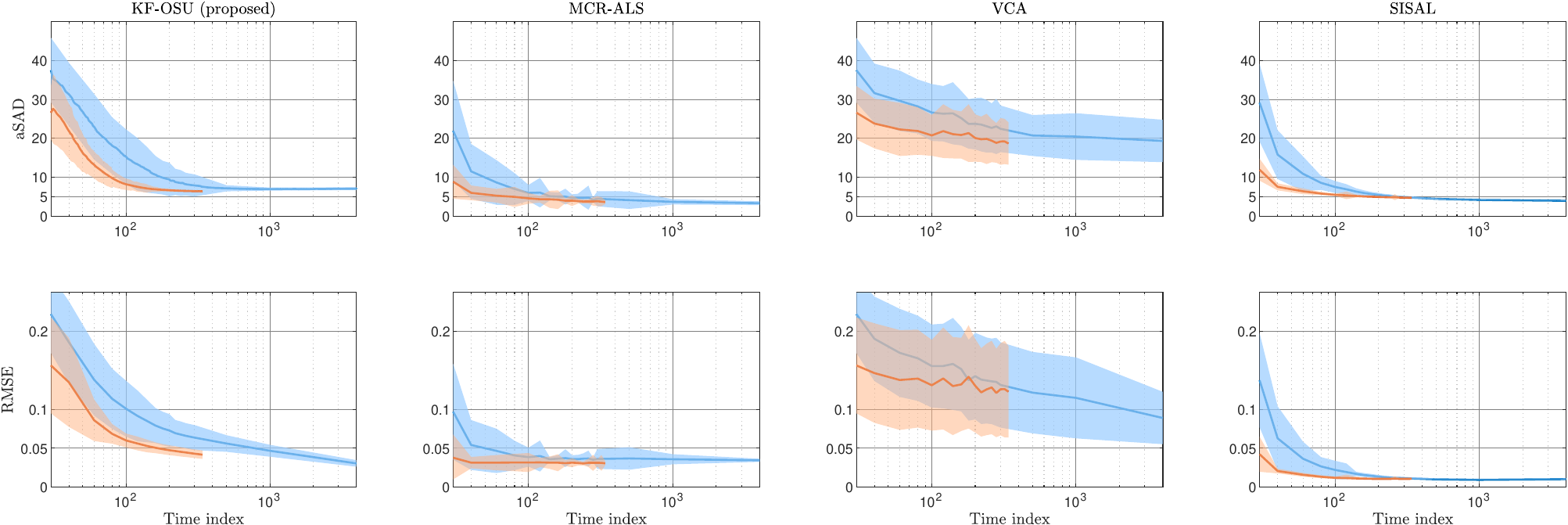}  \caption{Synthetic data set \texttt{SD2} -- Performance of the compared algorithms in terms of aSAD ($1$st row) and RMSE ($2$nd row) as functions of the time index. Blue and orange lines refer to the acquisition protocols \texttt{P1} and \texttt{P2}, respectively. The results have been averaged over $200$ data sets and the shaded areas correspond to one standard deviation.}
  \label{fig:fig2}
\end{figure*}

\subsection{Performance metrics}
Four metrics have been taken into account to evaluate the performance of the algorithms under study. The first one is the spectral angle distance (SAD, in degrees), defined by
\[ \text{SAD}_k = \arccos \left( \frac{\Hat{\bs}_k^\T\bs_k}{\left\| \Hat{\bs}_k \right\|_2 \left\| \bs_k \right\|_2} \right), \]
where \( \bs_k \) (resp. \(  \Hat{\bs}_k \) ) is the $k$th true (resp. estimated) PS. A small SAD value indicates a strong similarity between the profiles of the two spectra. To measure the average similarity over all the  PS, the average SAD (aSAD) is reported
\[ \text{aSAD} = \frac{1}{K}\sum_{k=1}^K  \text{SAD}_k.\]
Note that this metric is insensitive to multiplicative factors, i.e. it does not allow to assess whether the estimated PS are scaled versions of the true ones. To capture this information, the root mean square error (RMSE) of the concentrations is additionally calculated as
\[ \text{RMSE} = \sqrt{ \frac{1}{KN}\norm{  \bsC- \Hat{\bsC} }^2_\text{F}}, \]
where \(\bsC \) (resp. \( \Hat{\bsC}\)) is the true (resp. estimated) concentration matrix. Finally, to assess the ability of the compared methods to appropriately model the measurements, the spectral reconstruction error (RE), defined as
\[ \text{RE} = \frac{\norm{  \bsY- \Hat{\bsC}\Hat{\bsS}^\T }_\text{F}}{\norm{ \bsY }_\text{F}}\]
is also computed. The matrix \(\bsY \) contains the observed spectra and \( \Hat{\bsS}\) is the estimated PS matrix. In particular, RE is used to compare the performance of the methods when applied to the real data set \texttt{RD}, as the true PS and mixing concentrations are unknown. 

It is worth noting that these four figures-of-merits will be computed as functions of the so-called \emph{time index} of the compared unmixing algorithms. By this, one means that these metrics are computed after each new spectral pixel is acquired, in agreement with the considered operational context characterized by a sequential acquisition followed by an on-the-fly processing.

\begin{figure*}[t]
  \centering
\includegraphics[width=1\linewidth]{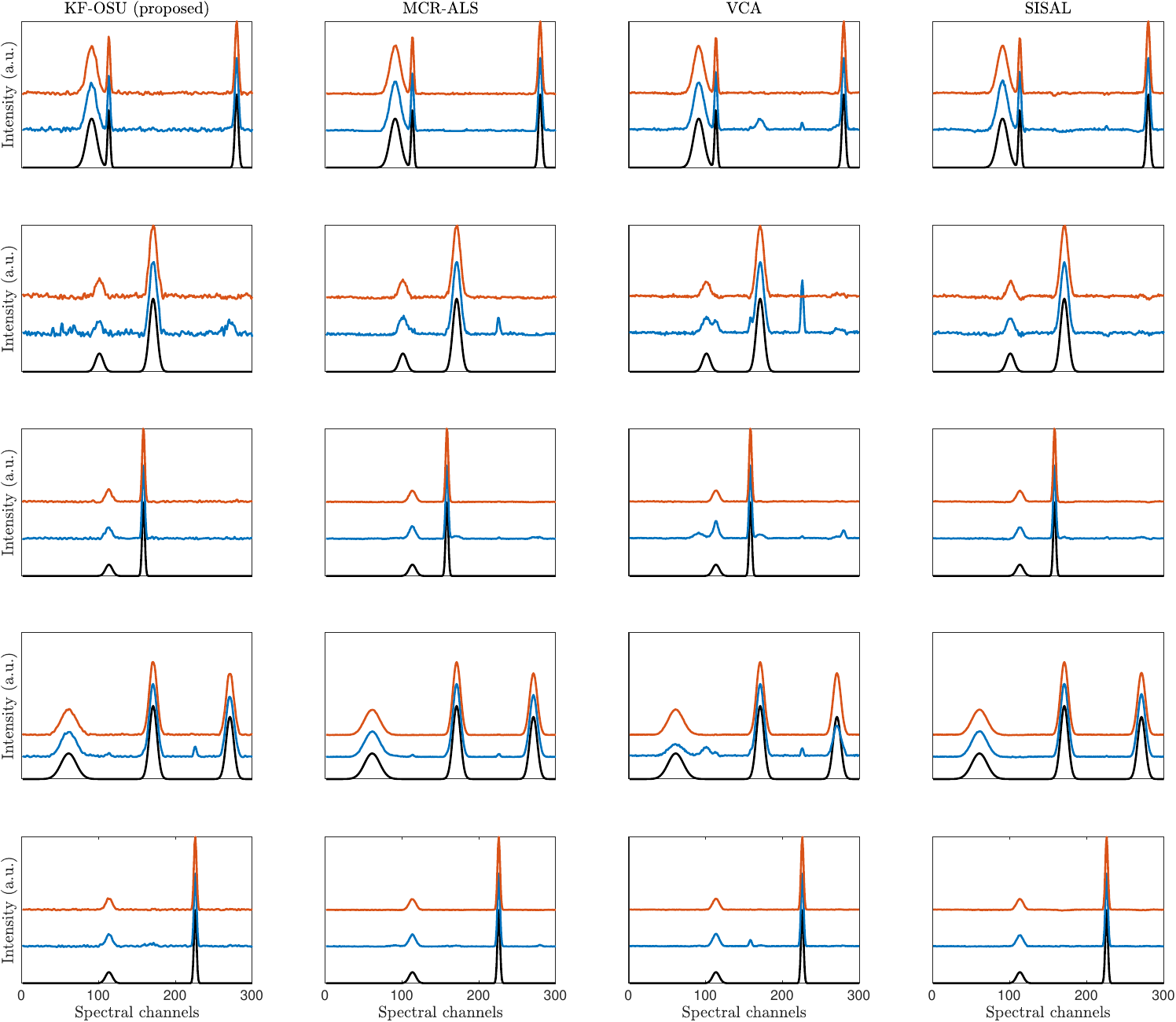}  \caption{Synthetic data set \texttt{SD1} -- Spectral profiles of the PS estimated at time index $t=200$ for acquisition protocols \texttt{P1} (blue) and \texttt{P2} (orange). The black curves correspond to the ground-truth spectra.}
  \label{fig:fig4}
\end{figure*}

\begin{figure*}[t]
  \centering
\includegraphics[width=1\linewidth]{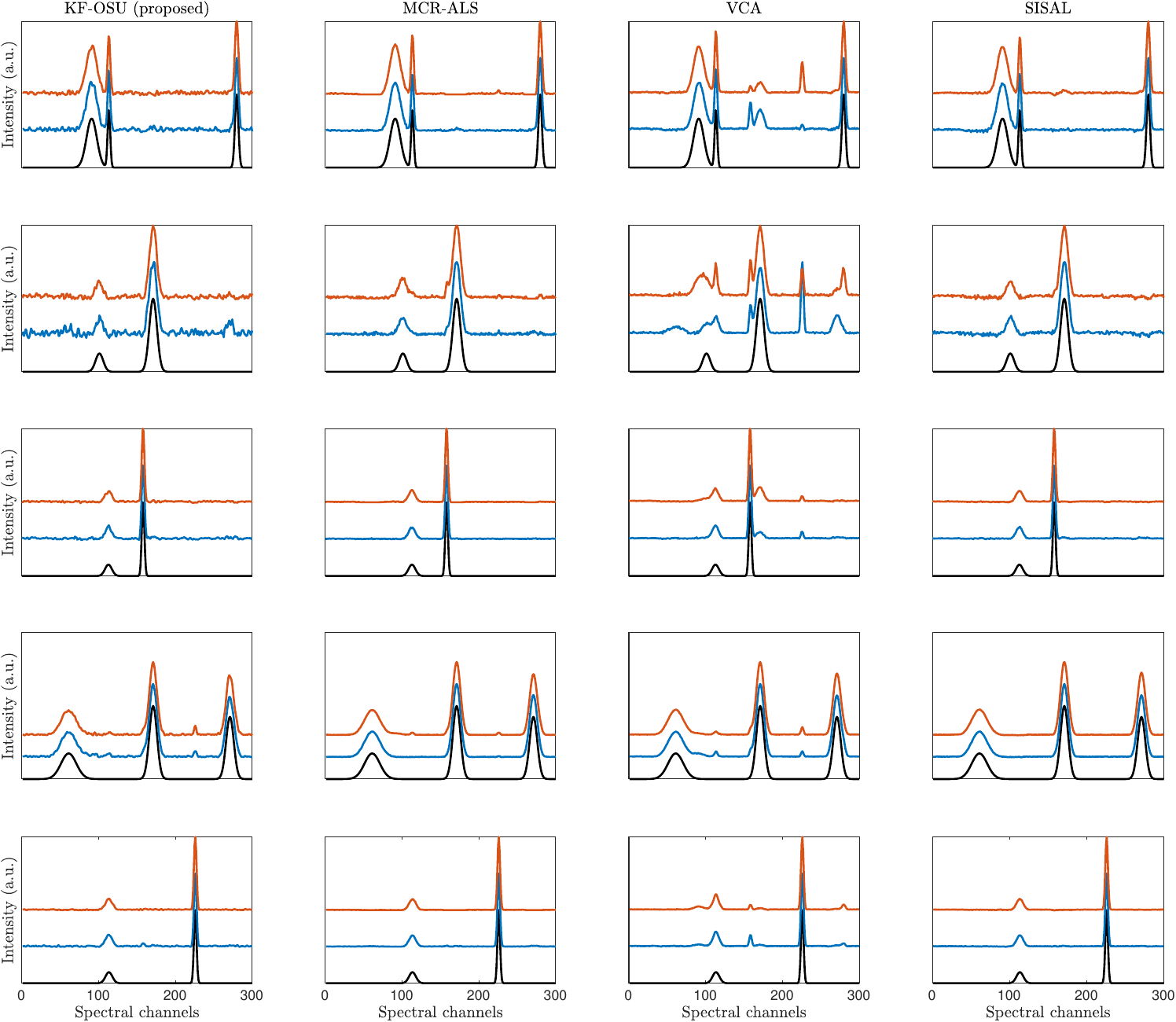}  \caption{Synthetic data set \texttt{SD2} -- Spectral profiles of the PS estimated at time index $t=200$ for acquisition protocols \texttt{P1} (blue) and \texttt{P2} (orange). The black curves correspond to the ground-truth spectra.}
  \label{fig:fig5}
\end{figure*}

\subsection{Tuning of KF-OSU parameters}
For each experiment conducted on the two synthetic data sets \texttt{SD1} and \texttt{SD2} and the real data set \texttt{RD}, the first $P=30$ spectra of each data set have been used to conduct the regression task (see~\eqref{eq11} and line~\ref{algostep:regression} of Algo. \ref{alg:KF-Algorithm}). 

The measurement noise level $\sigma^2_e$ has been estimated from these $P$ spectra after denoising using a Savitzky-Golay filter of order $3$ with a sliding window of size $5$  \cite{savitzky1964smoothing}. To avoid excessive under- or over-estimation of this noise level, the results have been averaged over several portions of spectrum. Formally, each of these $P$ spectra is split into $Q$ segments of size $T=10$ such that \(QT \approx L\). The variance $\sigma^2_e$ of the noise  is then set as the mean of the median variance calculated on each spectrum.

Given the inherent low-frequency nature of the Raman spectra used in these experiments, their representations in the lower-dimensionsal Fourier subspace have been computed by concatenating the real and imaginary parts of the first $M$ complex DFT coefficients (see \eqref{eq8}). This number $M$, estimated from these $P$ spectra, has been chosen as the minimum value for which at least \(\eta\%\) energy has been kept, i.e., such that the following criterion is met 
\[ 
\sum_{t=1}^{P} \norm{\mathsf{DR}\left({\bsy}_t\right)}^2_2
 \geq \frac{\eta}{100}\sum_{t=1}^{P} \norm{\bsy_t}^2_2.
\]
Table~\ref{tab:parameters} reports the (range of) values chosen for these parameters $\sigma^2_e$, $M$ and $\eta$ for each data set. It has been empirically observed that the uncertainty level \(\sigma^2_v \) should be greater than or equal to $1$; it has been set as $\sigma^2_v = 1$ for all experiments. Finally, the PS have been initialized as the outputs of VCA.  When solving the regression task (see Algo.~\ref{alg:ADMM}), the stopping criterion is chosen as a maximum of \(50\) iterations with \(\rho=1 \)  and \(\bsU^{(0)} \) and \(\bslambda^{(0)} \) initialized as null matrices. Protocol \texttt{P2} sketched in Algo. \ref{alg:P2} has been implemented  with $N_{\mathrm{ess}}=340$ for the data sets \texttt{SD1} and \texttt{SD2} and $ N_{\mathrm{ess}}=400$ for the real data set \texttt{RD}. The number of clusters used in protocol \texttt{P2} has been set to $J=50$ for all data sets.  

\begin{table}
    \centering
    \caption{Algorithmic parameter values for each data set.}
    \label{tab:parameters}
    \small
    \renewcommand{\arraystretch}{1.2} 
    \setlength{\tabcolsep}{8pt} 
    \begin{tabular}{ccccc}
    \toprule
    \multirow{2}{*}{\textbf{Data set}} & \multirow{2}{*}{{$M$}} & \multirow{2}{*}{{$\eta$ ($\%$)}} & {{$\hat\sigma^2_e$}} & {{$\sigma^2_e$}} \\ 
                               &              &              & (estimated) & (true value) \\ \midrule
    \multirow{1}{*}{\texttt{SD1}}        & 14--16                        & {85--90}  & 13          & 25     \\ 
    \multirow{1}{*}{\texttt{SD2}}        & 14--16                        & {85--90}  & 13           & 25         \\
    \multirow{1}{*}{\texttt{RD}}      & 22--25                        &{82--88}                 &48                & -               \\ \bottomrule
    \end{tabular}%
\end{table}

\subsection{Compared methods}
The proposed algorithm is compared to three conventional unmixing methods described below:
\begin{itemize}
    \item MCR-ALS is a popular unmixing algorithm in the chemometrics literature, thanks to its ease of implementation and its ability to produce good estimates of PS \cite{tauler1993multivariate}. MCR-ALS requires as input the number of PS, the maximum number of iterations (set to $60$ in this paper) and an initial estimate of  PS chosen as the outputs of PCA.
    \item VCA is a well-known geometric method widely used by the remote sensing community thanks to its computational efficiency and its ability to identify PS among observations when they have been measured \cite{nascimento2005vertex}. It is  considered in the comparison since it has been used to initialize KF-OSU.
    \item SISAL is another well-known algorithm widely used by the remote sensing community \cite{bioucas2009variable}. Contrary to VCA, it does not assume that PS have been measured. SISAL shows a competitive (often better) processing time than its peers and generally produces good PS estimates. Besides the number of PS, it requires as input the maximum number of constrained quadratic problems to be solved, set to $80$ in the experiments.
\end{itemize}
These algorithms are designed to process data in an offline context, i.e., to identify PS from a whole given set of measured spectra. To compare them to KF-OSU, they have been implemented to satisfy the operational requirements imposed by a sequential acquisition protocol combined with an on-the-fly spectral processing. More precisely, at each time instant $t$, these three algorithms are always run on the entire set of available acquired spectra  $\bsY_{\leq t}$. In other words, after each new spectrum $\bsy_t$ is acquired, MCR-ALS, VCA and SISAL provide a new PS estimate which is computed from all previous acquired spectra  $\bsY_{\leq t-1}$ augmented by this newly acquired spectrum $\bsy_t$. These three methods are expected to reach good estimation performance. However, it will be shown that their respective computational burden makes them inoperative in the targeted practical setup, imposing an on-the-fly unmixing of data acquired sequentially according to a spectrum-by-spectrum scheme.

\section{Results and discussion}\label{sec:discussion}

\subsection{Synthetic data sets}

Figure~\ref{fig:fig3} shows the performance of the compared methods on the data set \texttt{SD1}. Overall, all algorithms yield good results, characterized by decreasing aSAD, RMSE and uncertainty (i.e., standard deviation) along the time index. Although the proposed method performs slightly worse than MCR-ALS and SISAL, it presents fairly low aSAD and RMSE values from time index $t=500$ for protocol \texttt{P1} (without the selection of essential spectra) and from time index $t=200$ for protocol \texttt{P2} (with the selection of essential spectra). This performance discrepancy  with respect to the compared methods may be partly explained by the quite low SNR of the investigated data. Since {KF-OSU} operates in a purely on-the-fly context, a spectrum-by-spectrum processing may be less effective to mitigate the impact of noise. Conversely, the compared methods may benefit from the simultaneous of the whole data sets, since they can infer the signal subspace at each time instant before identifying the PS.

Besides, it can be observed in Fig.~\ref{fig:fig3} that the aSAD and RMSE curves of the four methods reach relatively close final values for the two acquisition protocols. However, it is important to highlight that under the second acquisition protocol \texttt{P2}, the aSAD and RMSE values decrease faster (in comparison with the experiment conducted under protocol \texttt{P1}) while the unmixing improves. This illustrates the benefit of resorting to such a smart acquisition protocol. In particular, one can notice that the curves associated to VCA reach a stationary behavior very late under protocol \texttt{P1}. This results from the fact that pure (or at least essential) spectra may be observed very late under protocol \texttt{P1}, in contrast to \texttt{P2} which mimics an acquisition sequence maximizing the spectral diversity. This shows once again the relevance of a smart acquisition protocol in the context of on-the-fly unmixing.

All previous findings are corroborated by the outcomes resulting from the analysis of the data set \texttt{SD2}, as shown in Fig.~\ref{fig:fig2}. The performance of {KF-OSU}, SISAL and MCR-ALS remains broadly the same as when the data sets \texttt{SD1} are concerned. This demonstrates their ability to provide good PS estimates even in the absence of pure pixels among the data. As expected, VCA shows poorer results, since this geometrical method explicitly assumes that the sought PS belong to the set of measured spectral pixels.

Figures~\ref{fig:fig4} and \ref{fig:fig5} display the profiles of the pure spectra estimated  by the compared methods at time instant $t=200$, i.e., after the $200$th acquisition. They show that the estimated PS signatures  are fairly close to the true PS. This demonstrates the ability of {KF-OSU} to perform on-the-fly unmixing without reprocessing all the previously measured spectra at each new acquisition.

\subsection{Real data set}

\begin{figure*}[t]
  \centering
\includegraphics[width=1\linewidth]{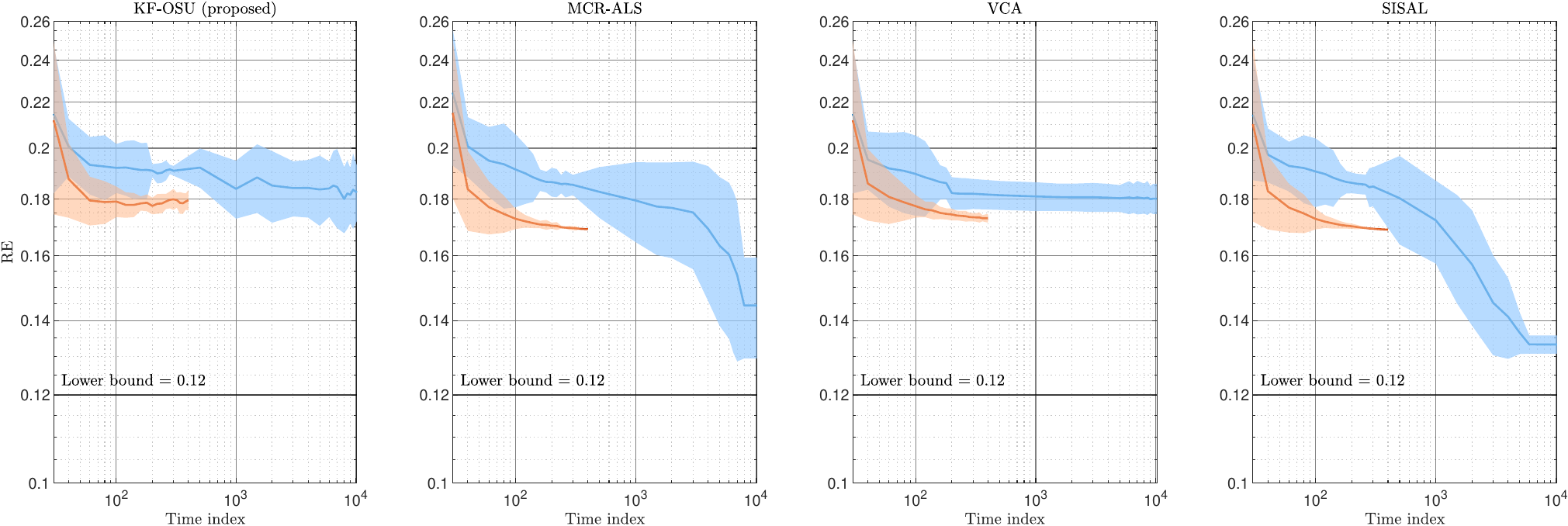}  \caption{Real data set \texttt{RD} -- Performance of the compared algorithms in terms of RE as a function of the time index. Blue and orange lines refer to the acquisition protocols \texttt{P1} and \texttt{P2}, respectively. The results have been averaged over $200$ data sets and the shaded areas correspond to one standard deviation.}   
  \label{fig:fig6}
\end{figure*}

Figure~\ref{fig:fig6} depicts the RE as a function of the time index for each of the compared methods. These curves are complemented with a horizontal line corresponding to a so-called \emph{lower bound} which approximates the lowest RE that can be expected given the noise level. This lower bound is defined as the RE with respect to $\bar{\bsY}$ resulting from a denoising by PCA, i.e., 
\begin{equation}
\mathrm{lower \ bound} = \frac{\norm{  \bsY- \bar{\bsY} }_\text{F}}{\norm{ \bsY }_\text{F}}    
\end{equation}
with $\bar{\bsY} = \bsY\bsP\bsP^\T$ where \(\bsP \in \R^{L \times K}\) is the matrix composed of the first $K$ principal component loadings ($K=3$). These curves show that all methods behave similarly for each protocol with a clear decreasing of the RE along the time index, i.e., when the number of measured spectra increases. Under protocol \texttt{P1}, MCR-ALS and SISAL reach significantly lower RE (close to the lower bound) when all the spectral pixels are considered to perform unmixing. Under this same protocol, KF-OSU and VCA provide higher RE, even when all spectral pixels have been considered. In general, the RE obtained under protocol \texttt{P2} are slightly smaller and, above all, they clearly decrease faster along the first $400$ observations, with less dispersion around the mean. 

It is worth noting that the slightly better results obtained by MCR-ALS and SISAL come with a higher computational burden, since all available spectral pixels are considered at each new measurement. This remains incompatible with the operational constraints imposed by a sequential acquisition protocol combined with an on-the-fly data processing, as it will be discussed in Section \ref{subsec:runtime}. In particular, they require a larger amount of spectra and, thus, a longer data collection time which, in the context of biological imaging, would significantly increase the risk of damaging the target specimens.

\begin{figure*}[t]
  \centering
\includegraphics[width=1\linewidth]{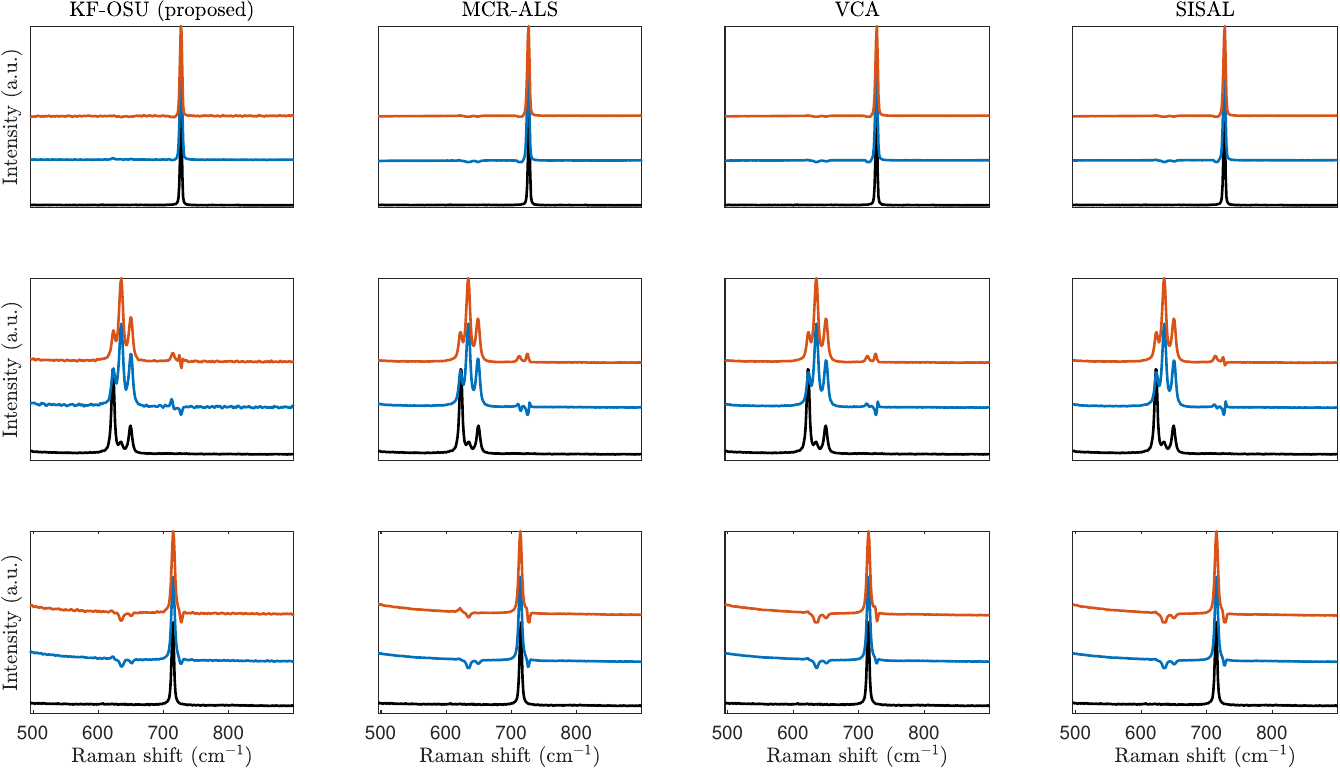}  \caption{Real data set \texttt{RD} -- Spectral profiles of the PS estimated at time index $t=140$ for acquisition protocols \texttt{P1} (blue) and \texttt{P2} (orange). The black curves correspond to the reference spectra.}
  \label{fig:fig7}
\end{figure*}

Finally, Fig.~\ref{fig:fig7} shows that the profiles of the PS estimated by the methods at time index $t=140$ can be easily matched to reference spectra associated to the main components of the sample. They are globally in good agreement with these spectra that can be roughly considered as ground truth. Note that under the simulated acquisition protocol \texttt{P1}, a higher reconstruction error between Raman shifts $700$ and $750$ cm$^{-1}$ is observed for the second PS.

\subsection{Runtime comparison}\label{subsec:runtime}

Table~\ref{tab:runtime_comparison} reports the computational times required by the compared algorithms for the processing of the synthetic data sets \texttt{SD1} and \texttt{SD2} and the real data set \texttt{RD}. It should be emphasized that KF-OSU updates the PS estimates at each time instant from a single newly available observation, i.e,  the last acquired spectrum. Consequently, it has a constant execution time per time index. On the contrary, other compared methods estimate the PS from the whole set of  spectra available at each time instant, which results in computational complexities that increase along the time index. As expected, KF-OSU is shown to be computationally more suitable for on-the-fly SU.

Furthermore,  the processing time  required by MCR-ALS for the $N=4000$ spectra of the data set \texttt{SD1} is much higher than the time required to analyze the data set \texttt{SD2}. This is certainly due to the absence of PS in the observations, resulting in a slower convergence of the algorithm. A similar observation can be made for SISAL, while KF-OSU takes the same execution time on both data sets. 

When considering the real data set \texttt{RD}, KF-OSU shows a lower execution time compared to when it was run on the synthetic data sets \texttt{SD1} and \texttt{SD2}. This can be explained by the number of pure spectra to be estimated ($K=3$), which is smaller than for the data sets \texttt{SD1} and \texttt{SD2} ($K=5$). This is not the case for MCR-ALS, VCA and SISAL, whose computational runtimes required to unmix the real data set \texttt{RD} are higher. In particular, VCA and SISAL execution times double with respect to the simulation experiments. This may be explained by the fact that their computational burden  is also highly driven by the difficulty of the SU task when analyzing real data.

\begin{table}
    \centering
    \caption{Average runtimes required by the compared SU algorithms. All the SU methodologies under study were} implemented using MATLAB R2022b and run on a laptop with an Intel Core i7-8565U CPU running at 1.80GHz equipped with 8.00GB RAM.
    \label{tab:runtime_comparison}
    \small
    \renewcommand{\arraystretch}{1.2} 
    \setlength{\tabcolsep}{4pt} 
    \resizebox{\columnwidth}{!}{
    \begin{tabular}{cccccc}
    \toprule
    \multirow{2}{*}{\textbf{Data set}} & \multirow{2}{*}{{$N$}} & \textbf{KF-OSU} & \textbf{MCR-ALS} & \textbf{VCA} & \textbf{SISAL}   \\ 
                               &                            & {(per obs.)} & (for $N$ obs.) & (for $N$ obs.) & (for $N$ obs.) \\ \midrule
    \multirow{1}{*}{\texttt{SD1}}        & 4000                        & {0.007}  & 1.38          & 0.04     & 0.15        \\ 
    \multirow{1}{*}{\texttt{SD2}}        & 4000                        & {0.007}  & 0.4           & 0.04     & 0.12         \\
    \multirow{1}{*}{\texttt{RD}}      & 10201                        &{0.005}                 &0.45                &0.08          &0.25               \\ \bottomrule
    \end{tabular}%
    }
\end{table}

\section{Conclusion}\label{sec:conclusions}
This work introduced a novel algorithm able to perform spectral unmixing in an on-the-fly setup. This method, named as KF-OSU, was specifically designed to be compatible with sequential acquisition protocols delivering measurements on a spectrum-by-spectrum basis. Indeed, KF-OSU was able to update the pure spectra estimates after each newly acquired individual spectrum. The evolution of the estimates as well as the spectral mixtures were described by linear Gaussian models, leading to simple and computationally efficient updating rules thanks to a Kalman filtering scheme. To further reduce the computational cost, a dimensionality reduction based on the discrete Fourier transform was performed beforehand. The nonnegativity of the pure spectra estimate was ensured by solving a regression problem relating the pure spectra in the original data space and their representations in the lower-dimensional Fourier subspace.

Experiments conducted on synthetic and real Raman data sets showed that the estimation performance of KF-OSU was comparable to those obtained by standard unmixing methods, which were developed to estimate pure spectra from a whole set of measurements. Moreover, an experimental analysis of the runtimes required by the compared methods demonstrated the computational efficiency of KF-OSU, whose cost was found to be in agreement with requirements imposed by a sequential data acquisition protocol combined with an on-the-fly spectral processing.

To the best of the authors' knowledge, this work is the first proposing a method able to perform  unmixing of spectral data delivered on a spectrum-by-spectrum basis. It should be a significant asset to support the development of guided image acquisition or smart scanning schemes  \cite{coic2023assessment,gilet2024superpixels}. 

\appendix
\section{Resolution of the regression problem}
\label{sec: appendix1}
This appendix provides practical details on the resolution of the regression task in \eqref{eq11}. This task is formulated as the minimization of a quadratic term subject to inequality constraints. After introducing a splitting variable \(  \bsU \in  \R_{+}^{L \times K}\), the problem can be equivalently rewritten as
\begin{equation}
    \min_{\bsR,\bsU} \norm {\Tilde{\bsY}_{1:P}\bsR - \htS_t}^2_\text{F} + \iota_{\mathbb{R}_+}(\bsU) 
    \quad \textrm{s.~t.} \quad \bsU - \bsY_{1:P}\bsR  = \bszero,
    \label{eq12}
\end{equation}
where $\iota_{\mathbb{R}_+}(\cdot)$ is the indicator function such that \( \iota_{\mathbb{R}_+}(\bsU)=0 \) if \( \bsU \succeq \bszero \) and  \( \iota_{\mathbb{R}_+}(\bsU) = \infty \) otherwise. The augmented Lagrangian associated to the constrained problem in \eqref{eq12} is
\begin{align*}
    \Lcal(\bsR, \bsU, \bslambda) =&  \norm{\Tilde{\bsY}_{1:P}\bsR - \htS_t}^2_\text{F} + \frac{\rho}{2}\norm{ \bsU - \bsY_{1:P}\bsR}^2_\text{F}\\
    &   + \langle \bslambda,  \bsU - \bsY_{1:P}\bsR \rangle + \iota(\bsU)
\end{align*}
where \(  \langle \bsX,  \bsZ\rangle = \tr{(\bsX^\T\bsZ)} \) and \( \bslambda \) is the Lagrange multiplier. Minimizing \eqref{eq12} can thus be efficiently achieved using ADMM whose scheme is detailed in Algo. \ref{alg:ADMM}. 
Solutions to lines \ref{admm_step:update_R} and \ref{admm_step:update_A} are given respectively by
\begin{align}
   \bsR^{(j+1)} =& \left( 2\Tilde{\bsY}_{1:P}^\T\Tilde{\bsY}_{1:P} + \rho\bsY_{1:P}^\T\bsY_{1:P}  \right)^{-1} \nonumber\\
    & \times \left(\bsY_{1:P}^\T\bslambda^{(j)} + \rho\bsY_{1:P}^\T\bsU^{(j)} + 2\Tilde{\bsY}_{1:P}^\T\htS_t \right) \label{eq:resolution_R}
\end{align}
and
\begin{equation}\label{eq:resolution_U}
    \bsU^{(j+1)} = \max\left(\bszero, \bsY_{1:P}\bsR^{(j+1)} - \tfrac{1}{\rho}\bslambda^{(j)}\right).
\end{equation}
It appears that the number $P$ of regressors should be chosen such that the matrix to be inverted in \eqref{eq:resolution_R} is well conditioned, or at least, nonsingular.

\begin{algorithm}[t]
\caption{\textsf{Regression}}\label{alg:ADMM}
\SetKwInOut{KwIn}{Input}
\SetKwInOut{KwOut}{Output}
    \KwIn {The regressors $\bsY_{1:P}$ in the original data space, their counterparts $\Tilde{\bsY}_{1:P}$ in the lower-dimensional Fourier subspace, the PS estimate $\htS_t$ expressed in the lower-dimensional Fourier subspace, the step-size $\rho>0$}
\SetKwInOut{KwIn}{Initialization}    
    \KwIn {\(\bsU^{(0)}, \bslambda^{(0)}\)}
    
    \( j \gets 0 \)\\
    \Repeat{\textnormal{stopping criterion}}{
        \mycomment{Update the regression matrix (see \eqref{eq:resolution_R})}
        \( \bsR^{(j+1)}  \in \operatornamewithlimits{argmin}_{\bsR} \Lcal(\bsR, \bsU^{(j)}, \bslambda^{(j)}) \) \label{admm_step:update_R}\\
        \mycomment{Update the splitting variable (see \eqref{eq:resolution_U})}
        \( \bsU^{(j+1)}  \in \operatornamewithlimits{argmin}_{\bsU} \Lcal(\bsR^{(j+1)}, \bsU, \bslambda^{(j)}) \) \label{admm_step:update_A}\\
        \mycomment{Update the Lagrange multiplier}
        \( \bslambda^{(j+1)} \gets \bslambda^{(j)} + \rho(\bsU^{(j+1)} - \bsY_{1:P}\bsR^{(j+1)}) \)\\
        \(j \gets j+1 \)\\
    }
    \KwOut {The regression matrix \( \bsR^{(j)} \). }
    \end{algorithm}

\section{Algorithmic sketch of the acquisition protocol \texttt{P2}}
\label{sec: appendix2}
The algorithmic sketch of the simulated acquisition protocol \texttt{P2} is given in Algo.~\ref{alg:P2}. 
Candidate spectral pixels are first identified by successive convex hull calculations and subsequently stored in the matrix $\bsZ$ (line \ref{line:convhull_calculation}), following a peeling process (line \ref{line:peeling}) similar to the one adopted in \cite{ahmad2023weighted}. These pixels are then partitioned into $J$ classes using \textsf{K-means} and a representative (centroid) of each class is then identified, denoted as $\bsm_j$, $j=1,\dots,J$ (line \ref{line:Kmeans}). Finally, essential spectral pixels are selected as the measured spectra which are closest to the centroids (line \ref{line:ESP_identification}) and their observation indices are stored in $\mathcal{J}$ (line \ref{line:index_storage}). A permutation $\sigma(\cdot)$ is finally introduced into $\mathcal{J}$ to produce different order sequences (line \ref{line:permutation}).

\begin{algorithm}[p]
\caption{Acquisition protocol \texttt{P2}}\label{alg:P2}
\SetKwInOut{KwIn}{Input}
\SetKwInOut{KwOut}{Output}

\KwIn{Data set \(\bsY\), number of essential spectra \(N_\text{ess}\), number of clusters $J$}

    \( \bsD \gets [] \)\\
    \( \bsZ \gets [] \)\\
    \( \bsW \gets \bsY \)\\
    \mycomment{Progressively extract successive convex hulls}
    \While{ \( \mathsf{card}(\bsZ) \, <  \) \(N_\mathrm{ess}\) }{    
    \(\bsZ  \gets \left[\bsZ, \mathsf{convhull}(\bsW)\right] \)  \label{line:convhull_calculation}\\
      \( \bsW \gets \bsY \setminus \bsZ \) \label{line:peeling}\\
    }
        \mycomment{Partition \(\bsZ\) into \(J\) clusters and identify the centroids}
        $\left\{\bsm_1,\ldots,\bsm_J\right\} \gets \textsf{K-means}(\bsZ,J)$ \label{line:Kmeans}\\
        \( n \gets 0 \) \\
         \( \bsW \gets \bsZ \)\\
        \While{\(n \le \) \(N_\mathrm{ess}\)}{
            \mycomment{Identify $J$ essential spectra (the closest to each centroid)}
            \( \mathcal{J} \gets [] \)\\
            \For{$j=1,\ldots,J$}
            {
            $k^\star \gets \operatorname{argmin}_{k} \left\|\bsm_j - \bsw_k\right\|^2_2$  \label{line:ESP_identification}\\
            $\mathcal{J} \gets \mathcal{J} \cup \left\{k^*\right\}$ \label{line:index_storage}
            }
            \mycomment{Randomly permute the indices in \(\mathcal{J} \)}
            $\check{\mathcal{J}} \gets \sigma\left(\mathcal{J}\right)$  \label{line:permutation}\\
            \mycomment {Complement the matrix of essential spectra}
            \( \bsD \gets [\bsD,\bsZ_{\check{\mathcal{J}}}] \)\\
            \mycomment {{ Discard \(\bsD \) from \(\bsZ \)}}
            \( \bsW \gets \bsZ \setminus \bsD \)\\
            \( n \gets n + J \)\\ 
        }
         \( \bsD \gets \bsD_{1:N_\text{ess}} \)\\
    \KwOut { The matrix of essential spectra \( \bsD \). }
\end{algorithm}

\bibliographystyle{elsarticle-num} 
\bibliography{biblio}

\end{document}